\documentclass[conference]{IEEEtran}
\IEEEoverridecommandlockouts

\usepackage{cite}
\usepackage{amsmath,amssymb,amsfonts}
\usepackage{algorithmic}
\usepackage{graphicx}
\usepackage{textcomp}
\usepackage{xcolor}
\usepackage{multirow}
\usepackage{array}
\usepackage{tikz} 
\usepackage{comment}
\def\BibTeX{{\rm B\kern-.05em{\sc i\kern-.025em b}\kern-.08em
    T\kern-.1667em\lower.7ex\hbox{E}\kern-.125emX}}

\usepackage{tikz} \usepackage{lipsum} \makeatletter \def\ieeecopyright{ \footnotesize © 2026 IEEE. Personal use of this material is permitted.\newline DOI: 10.1109/CCNC} \makeatother \AddToHook{shipout/firstpage}{\begin{tikzpicture}[remember picture,overlay] \node[anchor=south west,xshift=1.0cm,yshift=0.8cm] at (current page.south west){\parbox{\linewidth}{\raggedright\ieeecopyright}}; \end{tikzpicture} }
    
\begin{document}

\title{SC-MII: Infrastructure LiDAR-based 3D Object Detection on Edge Devices for Split Computing with Multiple Intermediate Outputs Integration\\
\thanks{This work was supported by JSPS KAKENHI Grant Number 23H00464.}
}

\author{
\IEEEauthorblockN{Taisuke Noguchi}
\IEEEauthorblockA{\textit{Graduate School of Science and Engineering} \\
\textit{Saitama University}\\
Saitama, Japan \\
t.noguchi.471@ms.saitama-u.ac.jp}
\vspace{-2mm}
\and
\IEEEauthorblockN{Takayuki Nishio}
\IEEEauthorblockA{\textit{School of Engineering} \\
\textit{Institute of Science Tokyo}\\
Tokyo, Japan \\
}
\vspace{-2mm}
\and
\IEEEauthorblockN{Takuya Azumi}
\IEEEauthorblockA{\textit{Academic Association} \\
\textit{(Graduate School of Science and Engineering)} \\
\textit{Saitama University}\\
Saitama, Japan \\
}
\vspace{-2mm}
}

\maketitle

\begin{abstract}
3D object detection using LiDAR-based point cloud data and deep neural networks is essential in autonomous driving technology.
However, deploying state-of-the-art models on edge devices present challenges due to high computational demands and energy consumption.
Additionally, single LiDAR setups suffer from blind spots.
This paper proposes SC-MII, multiple infrastructure LiDAR-based 3D object detection on edge devices for Split Computing with Multiple Intermediate outputs Integration.
In SC-MII, edge devices process local point clouds through the initial DNN layers and send intermediate outputs to an edge server.
The server integrates these features and completes inference, reducing both latency and device load while improving privacy.
Experimental results on a real-world dataset show a 2.19× speed-up and a 71.6\% reduction in edge device processing time, with at most a 1.09\% drop in accuracy.
\end{abstract}


\section{Introduction}
Autonomous driving technology~\cite{autoware} has advanced rapidly, offering significant societal benefits.  
A core component is object detection, which is essential for interpreting the environment and enabling route planning and vehicle control.  
While primarily based on vehicle-mounted sensors, infrastructure sensors can extend coverage, especially in blind spots.  
Importantly, LiDAR is a central sensor used in both vehicle and infrastructure setups~\cite{roslite}.

LiDAR captures 3D point clouds, which Deep Neural Networks (DNNs) analyze for object detection~\cite{deng2021voxel, li2023lightweight}.  
Infrastructure-based DNN inference is typically performed on edge devices.  
To improve detection reliability, multiple LiDARs can be used, but transmitting raw point clouds to an edge server raises privacy concerns.
Local inference reduces transmission needs, but edge devices often lack the power to run state-of-the-art models, leading to latency and power inefficiencies~\cite{li2023lightweight}.

This study addresses these challenges by enabling 3D object detection using multiple infrastructure LiDARs while minimizing processing time, edge-side load, and privacy risks. 
This paper adopts Split Computing (SC)~\cite{SC_survey, noguchi20243d}, where a DNN is split between edge devices and an edge server.  
Edge devices process early layers and send intermediate outputs to the edge server for final inference, reducing the computational burden on edge devices and avoiding raw data transmission.

To overcome these challenges, this paper proposes SC-MII, an infrastructure LiDAR-based 3D object detection method for SC with Multiple Intermediate outputs Integration.
Each edge device processes its local LiDAR data through DNN initial layers and transmits intermediate features to an edge server.  
Upon reception, the edge server spatially aligns these intermediate outputs, which retain position information from the original point clouds, to a common reference frame, integrates them, and completes inference.
The contributions of this paper are as follows:
\begin{itemize}
\item
Significant reduction in inference time by offloading computation from edge devices to an edge server.
\item
Lower computational burden on edge devices through partial execution of the DNN model.
\item
Enhanced detection accuracy compared to using a single LiDAR, achieved by integrating intermediate outputs from multiple LiDARs.
\end{itemize}

\section{System Model} \label{system_model}

\begin{figure}[t]
        \centerline{\includegraphics[width=1.0\linewidth]{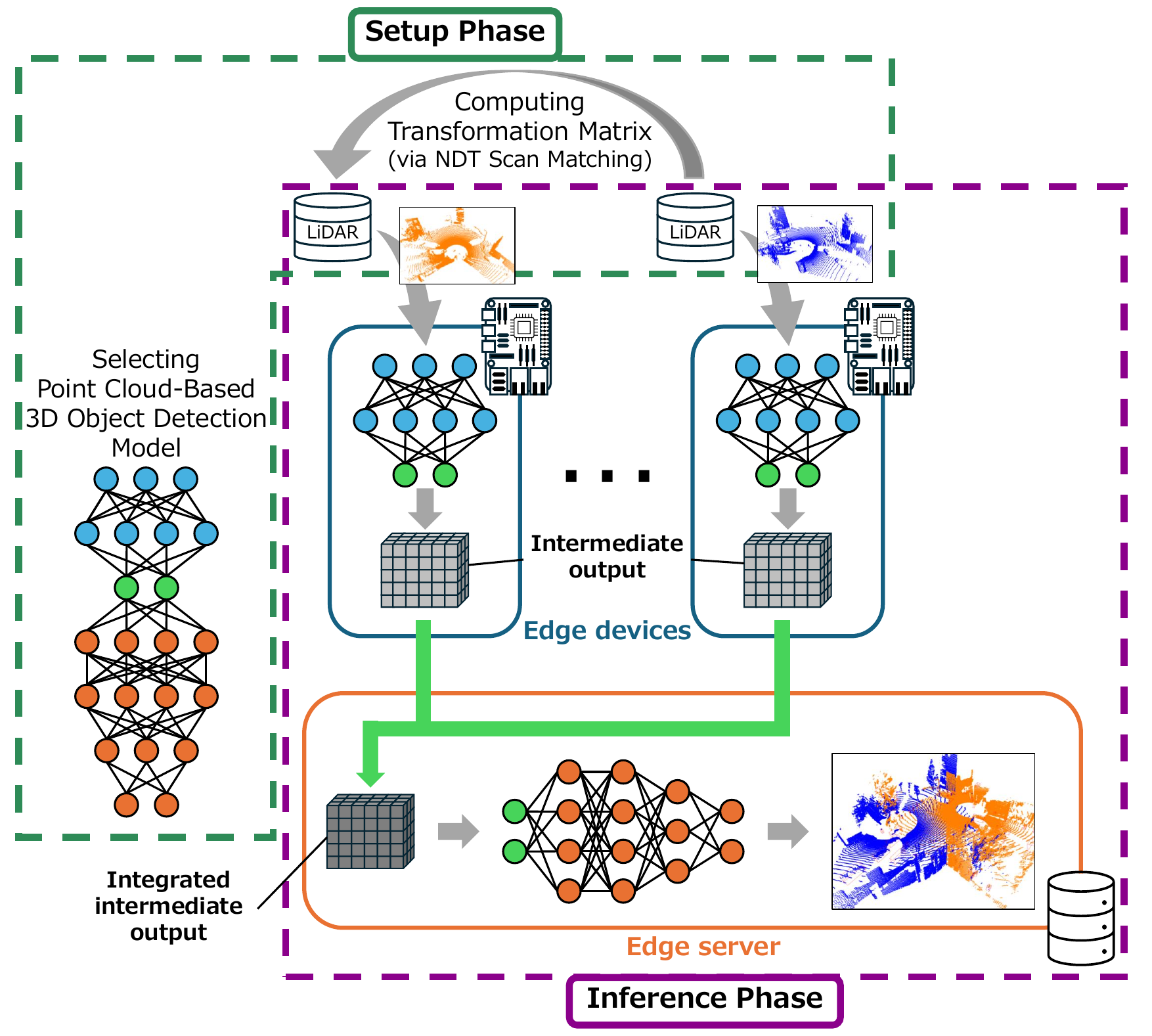}}
        \caption{System model.}
        \vspace{-5mm}
    \label{fig-system_model}
\end{figure}

This paper proposes a 3D object detection method using point cloud data from multiple LiDARs.
The system model of SC-MII is shown in Fig.~\ref{fig-system_model}.
The inference method is based on the SC idea.
In the setup, NDT Scan Matching is used to calculate the relative position of each LiDAR.

This section provides background relevant to the proposed method.
Section~\ref{split_computing} describes Split Computing, Section~\ref{object_detection} explains point cloud-based 3D object detection, and Section~\ref{ndt} discusses NDT Scan Matching.

\subsection{Split Computing (SC)} \label{split_computing}

SC is a method in which a DNN is split into two parts, executed separately on an edge device and an edge server.  
This approach reduces the computational load on the edge device while maintaining high accuracy and efficiency.  
The pre-trained model is split into a head model assigned to the edge and a tail model assigned to the server.
During inference, the edge device runs the head model on input data and transmits the intermediate output to the edge server, which runs the tail model and returns the final prediction.

SC optimizes processing by offloading calculations to the edge server, eliminating the need for lightweight models.
Placing splitting points early on further reduces edge-side load, power consumption, and inference time.

\subsection{Point Cloud-Based 3D Object Detection} \label{object_detection}

3D object detection is a key task in autonomous driving technology.  
Among various data formats, point clouds offer rich spatial information for accurate position estimation.  
They are commonly obtained from LiDAR sensors.
Point cloud-based detection methods are classified into the following three types.
Point-based methods~\cite{shi2019pointrcnn} directly process raw point clouds, while voxel-based methods~\cite{deng2021voxel} convert them into 3D grids called voxels for structured processing.
BEV-based methods~\cite{lang2019pointpillars} project point clouds onto a 2D plane, known as a bird’s-eye view, and then process them.
Object detection models also exist that combine these approaches~\cite{shi2020pv}.

\subsection{NDT Scan Matching} \label{ndt}

NDT Scan Matching is a method for simultaneous localization and mapping using point cloud data from LiDAR and other sensors.  
Normal Distributions Transform (NDT)~\cite{biber2003normal} models local regions of point clouds as normal distributions by dividing 3D space into voxels and computing the mean and covariance of each.
This yields a continuous, differentiable probability density function.

After generating the NDT representation for the reference point cloud, the current point cloud of the sensor is aligned by rotating and translating it to best match the reference.
Once aligned, the position and orientation of the sensor are estimated relative to the reference coordinate system.

\section{Approach} \label{approach}

A 3D object detection method is proposed that integrates information from multiple infrastructure LiDARs to address blind spots and improve detection accuracy.
One approach is to merge raw point cloud data from each LiDAR~\cite{chen2019cooper}.
However, this raises privacy concerns due to the need to transmit raw data to the server. 
Another method aggregates detection results independently on edge devices~\cite{xu2023model}, but edge devices with limited capacity struggle to run high-accuracy models without increased processing time or power consumption.
Lightweight models reduce resource usage but often compromise accuracy.

To overcome these limitations, this study adopts an approach that integrates intermediate outputs.
Edge devices compute the initial DNN layers and transmit intermediate outputs, rather than raw point clouds, to the edge server, where the features are aligned and integrated.
This avoids transmitting sensitive raw data and reduces the computational load on edge devices.

Our proposed method, SC-MII, performs 3D object detection by integrating intermediate features from multiple infrastructure LiDARs using the idea of split computing.
SC-MII consists of a setup phase and an inference phase.
In the setup phase, a transformation matrix is computed to align the coordinate systems of multiple infrastructure LiDARs, and the detection model is trained.
In the inference phase, each edge device computes intermediate outputs from its LiDAR input, which are transmitted to the edge server, aligned, integrated, and processed by the edge server to complete object detection.

The remainder of this section is organized as follows.
Section~\ref{inference_phase} describes the inference phase, including the inference flow, the coordinate transformation of intermediate outputs, and their integration.
Section~\ref{setup_phase} explains the setup phase, which covers transformation matrix computation, model selection and splitting, and model training.

\subsection{Inference Phase} \label{inference_phase}

\subsubsection{Inference Flow} \label{inference_flow}
The proposed method performs 3D object detection by integrating point cloud data from multiple infrastructure LiDARs.
In this setup, one edge server and as many edge devices as LiDARs are used.
Each edge device is paired with a LiDAR and continuously receives its point cloud data.
The object detection model, configured during the setup phase, is split into a head model and a tail model, assigned to the edge devices and the edge server, respectively.
All edge devices execute head models with the same architecture, but the parameters differ due to variations in the input data.

\begin{figure}[t]
    \centerline{\includegraphics[width=1.0\linewidth]{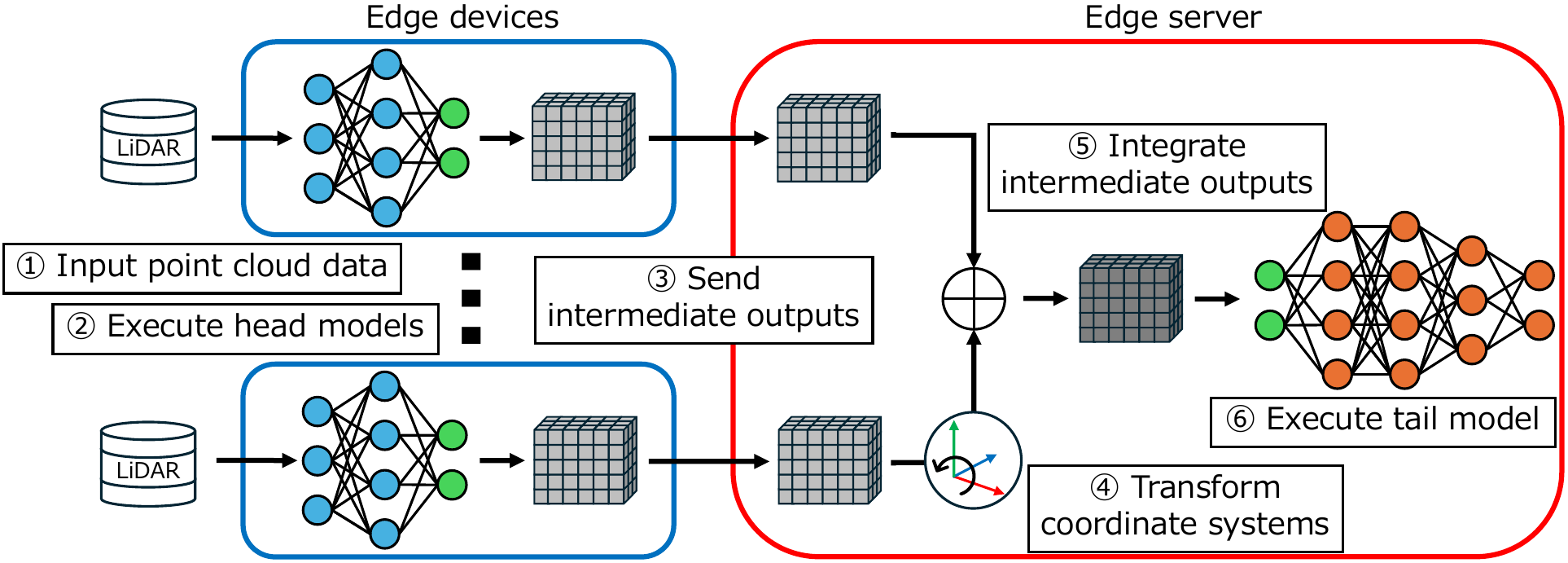}}
        \caption{Inference flow of SC-MII.}
        \vspace{-5mm}
    \label{fig-inference_phase}
\end{figure}

The inference flow of SC-MII is shown in Fig.~\ref{fig-inference_phase}.
Each edge device receives point cloud data from its corresponding LiDAR and runs the head model.
The resulting intermediate outputs are sent to the edge server, where they are aligned and integrated.
The edge server then runs the tail model using the integrated data to produce the final detection result, including object location and category.

\subsubsection{Coordinate Transformation of Intermediate Outputs} \label{transformation}
We assume that all edge devices generate intermediate feature maps on the same voxel grid resolution and with a common coordinate origin, ensuring that the transformation matrix can be applied directly to the feature tensors.
To spatially align the intermediate outputs from multiple edge devices into a common coordinate frame, a coordinate transformation is applied to the voxel-wise features obtained after sparse convolution.
Unlike conventional approaches that align raw point clouds before voxelization, this method transforms the intermediate feature maps produced after partial inference on the edge devices.

First, the discrete voxel indices of the intermediate outputs are converted into continuous physical coordinates.
This conversion is performed by scaling the voxel indices by the effective voxel size, which accounts for both the original voxel resolution and the scaling factor induced by convolutional operations, such as strided convolution, that alter the spatial resolution of the intermediate features.
The scaled indices are then shifted by the origin of the local coordinate system of each LiDAR sensor to obtain the physical coordinates.
Although the resulting physical coordinates remain aligned with the voxel grid, they are treated as continuous representations in the space for the purpose of coordinate transformation.

Subsequently, a known transformation matrix, estimated during the setup phase, is applied to the physical coordinates.
This matrix represents a rigid-body transformation that aligns the local coordinate system of each sensor with the common global reference frame.
The transformation is performed in homogeneous coordinates to incorporate both rotation and translation.

After applying the transformation, the physical coordinates are converted back into voxel indices by reversing the scaling and offset operations.
The resulting indices are rounded to the nearest integers on the voxel grid and constrained within the integration range.
This process enables spatially consistent integration of intermediate features from multiple sensors while preserving spatial locality.
This approach is feasible because convolutional operations, despite altering spatial resolution, preserve the spatial relationships of features in voxel space.

\subsubsection{Integration of Intermediate Outputs} \label{integration}
After the coordinate transformation, the intermediate outputs from multiple edge devices are spatially aligned within the common reference frame.
To integrate these aligned outputs, one of two methods is employed.

The first method selects the maximum value across all intermediate outputs at each voxel coordinate.
This approach requires only simple element-wise comparisons.
The second method concatenates all intermediate outputs along the feature dimension and applies a single convolutional layer to the concatenated features.
The output of this operation is designed to match the size of the original intermediate output from each device.
In this study, convolutional layers with kernel sizes of 1 or 3 are used for this integration method.
Both integration methods are evaluated for their effectiveness in downstream object detection tasks.
The methods are designed to be general and applicable to a wide range of object detection models.

\subsection{Setup Phase} \label{setup_phase}

\begin{figure*}[t]
    \centerline{\includegraphics[width=0.86\linewidth]{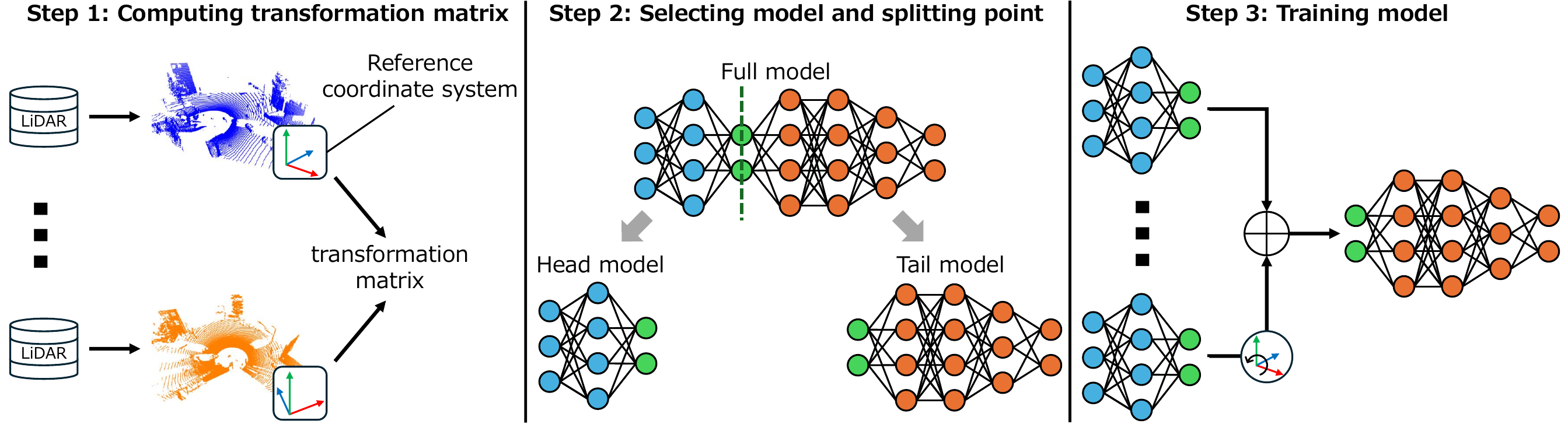}}
        \caption{Setup phase of SC-MII.}
        \vspace{-6mm}
    \label{fig-setup_phase}
\end{figure*}

The setup phase is conducted prior to system deployment and inference.
This process is shown in Fig.~\ref{fig-setup_phase}.

\subsubsection{Transformation Matrix Computation} \label{matrix_calc}

In SC-MII, intermediate outputs from multiple edge devices must be integrated in a common coordinate system.
Since each infrastructure LiDAR operates in its own local coordinate frame, it is necessary to compute transformation matrices that align these frames to a common reference frame.

\begin{figure}[t]
        \centerline{\includegraphics[width=0.89\linewidth]{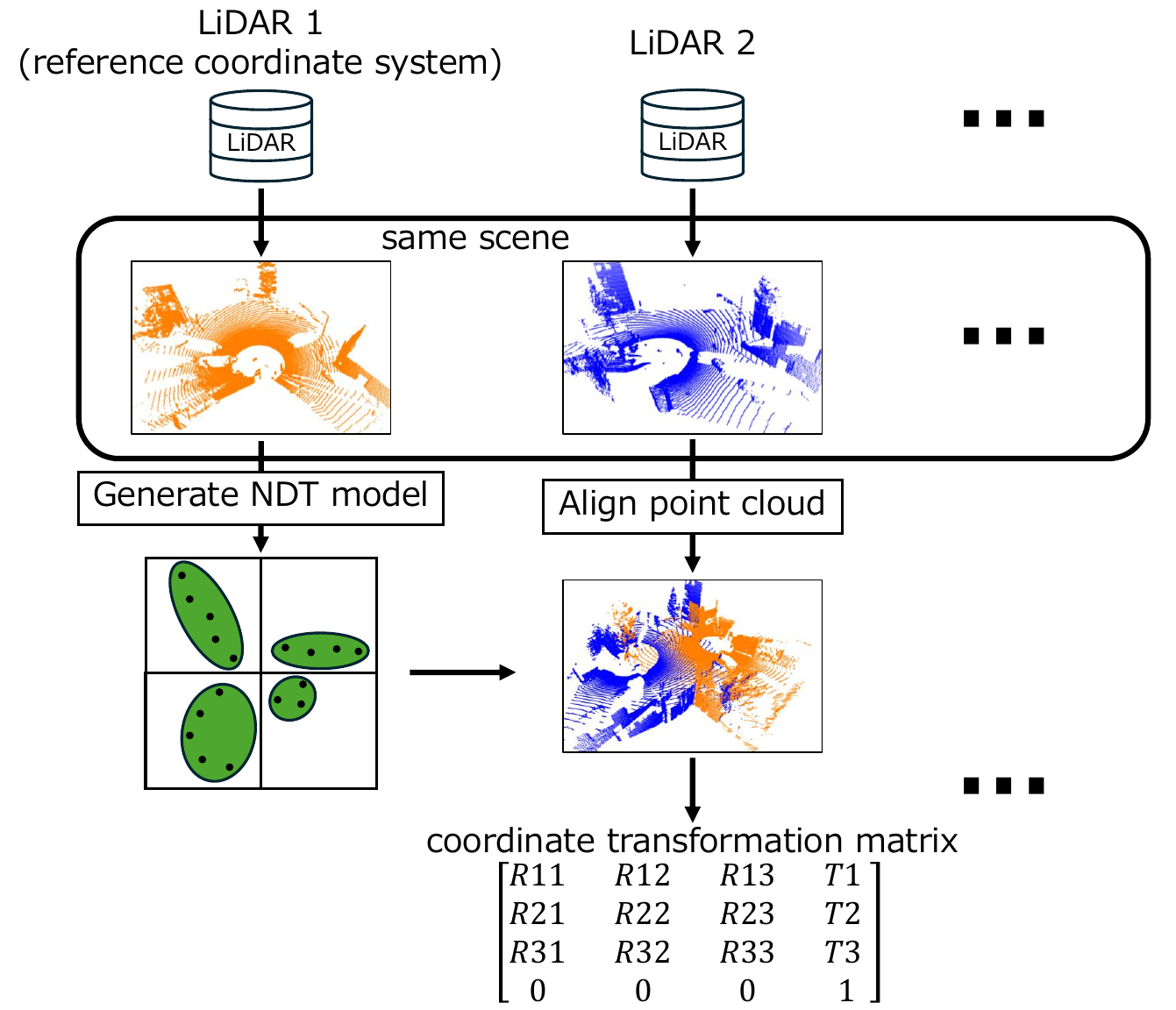}}
        \caption{Coordinate transformation matrix calculation via NDT scan matching.}
        \vspace{-5mm}
    \label{fig-trans_matrix}
\end{figure}

To estimate these transformation matrices, this study employs NDT scan matching. As illustrated in Fig.~\ref{fig-trans_matrix}, point cloud data from each LiDAR observing the same scene are collected beforehand.
One LiDAR is selected as the reference, and its point cloud is modeled using NDT, which represents local point distributions as normal distributions.
Point clouds from the other LiDARs are aligned to this model, resulting in a rigid-body transformation matrix for each sensor.

Because infrastructure LiDARs are fixed in position, unlike vehicle-mounted sensors, these transformation matrices remain valid over time.
The edge server stores these matrices and applies them as part of the coordinate transformation process, which converts intermediate outputs to a common coordinate frame and ensures spatial consistency across sensors.

\subsubsection{Model Selection and Splitting Point Strategies}

Model splitting is a critical design decision in SC-MII to balance communication efficiency, privacy preservation, and computational load distribution.
The splitting point within the 3D object detection model must satisfy two important constraints.

First, to minimize transmission time and bandwidth, the splitting point should be located at a layer producing small intermediate outputs.
Transmitting compact features reduces communication overhead and accelerates inference.
Branching structures in the model must also be considered, as they can increase output size due to additional data.
Second, the split must occur after sufficient preprocessing to avoid transmitting privacy-sensitive raw point cloud data.
This ensures that edge devices handle all raw data, protecting user privacy.
Models requiring raw point clouds in intermediate layers are unsuitable for SC-MII, as they would necessitate transmitting sensitive data or reduce the benefits of computational offloading.

Although optimization of model architecture and splitting points is an important research topic, this study adopts the configurations proposed in Ref.~\cite{noguchi20243d} and enhances them by merging compact intermediate feature maps from multiple infrastructure LiDARs within the network, thereby improving privacy preservation and reducing communication overhead.
Details of the specific models and splitting points used in our experiments are provided in Section~\ref{select_model_point}.

\subsubsection{Training Object Detection Models} \label{training}

Training is conducted offline during the setup phase prior to system deployment.
Prior to training, point cloud data collected from each LiDAR sensor, along with their corresponding labels, are aggregated and temporally synchronized.
The model architecture used for training is the same as that employed during inference, maintaining consistency between the two processes.
While inference involves distributed processing split between edge devices and the edge server, training is performed in a centralized, batch manner.
The raw point clouds themselves are not spatially transformed prior to training; rather, coordinate transformations are performed within the model on the intermediate features, consistent with the inference process.

\section{Evaluation} \label{evaluation}

This section evaluates SC-MII.
Each integration method is compared in terms of both accuracy and execution time, and trade-offs are investigated.

The remainder of this section is organized as follows.
Section~\ref{environment} provides an overview of the dataset utilized for evaluation, as well as the environment and equipment used during the evaluation process.
Section~\ref{select_model_point} discusses the selection of the model to be applied in SC-MII, along with the determination of splitting points.
Section~\ref{accuracy} evaluates the accuracy of each proposed integration method.
Section~\ref{execution_time} examines the execution time of each integration method.

\subsection{Dataset and Evaluation Environment} \label{environment}

This study uses the V2X-Real dataset~\cite{xiang2025v2x}, a real-world dataset for Vehicle-to-Everything (V2X) cooperative perception.  
This dataset was collected at an intersection using two infrastructure sensors and two Connected Autonomous Vehicles (CAVs), each equipped with a LiDAR and a camera.
This study uses two infrastructure LiDARs: Ouster OS1-64 and OS1-128, both operating at 10 Hz.

\begin{table}[t]
    \centering
    \caption{Evaluation Environment}
    \scalebox{0.95}{
    \begin{tabular}{ll}
         \hline
         Components & Specifications \\ \hline
         \multirow{5}{*}{Edge devices} & Jetson Orin Nano: \\ 
         \ & - hexa-core Arm 1.5 GHz CPU \\
         \ & - 1024-core NVIDIA Ampere architecture GPU \\
         \ & \quad with 32 Tensor Cores operating at 625 MHz \\
         \ & - 8 GB of 128-bit LPDDR5 memory \\ \hline
         \multirow{3}{*}{Edge server} & Intel\textregistered~Core i9-14900K CPU \\ 
         \ & NVIDIA RTX 4090 GPU \\
         \ & 192 GB of DDR5 memory \\ \hline
         Communication & 1 Gbps wired LAN \\ \hline
    \end{tabular}
    }
    \vspace{-1mm}
    \label{tab-environment}
\end{table}

\begin{table}[t]
    \centering
    \caption{Assignment of Infrastructure LiDARs and Corresponding Sensor Models for Each Edge Device}
    \scalebox{1.0}{
    \begin{tabular}{ccc}
         \hline
         Device & Assigned LiDAR & LiDAR model \\ \hline
         Device 1 & LiDAR 1 & Ouster OS1-64 \\
         Device 2 & LiDAR 2 & Ouster OS1-128 \\ \hline
    \end{tabular}
    }
    \vspace{-4mm}
    \label{tab-device_LiDAR}
\end{table}

In SC-MII, processing is split between edge devices and an edge server.  
This study uses two Jetson Orin Nano devices as edge devices and a high-performance server.  
Hardware and communication details are summarized in Table~\ref{tab-environment}.
Device 1 is assigned OS1-64, and Device 2 is assigned OS1-128, as shown in Table~\ref{tab-device_LiDAR}.  
Due to the sensor specifications, Device 2 processes roughly twice the number of points as Device 1.  

\subsection{Model Selection and Determination of Splitting Points} \label{select_model_point}

To ensure privacy and reduce transmission size, SC-MII requires a 3D object detection model that uses point cloud data only in its initial stage.
A prior study applying SC to 3D object detection~\cite{noguchi20243d} analyzed two models, PV-RCNN~\cite{shi2020pv} and Voxel R-CNN~\cite{deng2021voxel}, based on their module architectures and execution time distributions.

PV-RCNN is unsuitable for SC because it reuses raw point cloud data in later layers, increasing the amount of data that must be transmitted and raising privacy concerns.
In contrast, Voxel R-CNN accesses the point cloud only during the initial voxelization step.
The following layers operate solely on voxelized features, making it a better fit for SC-MII.

The same study identified the most appropriate splitting point by considering the trade-off between privacy and execution time.
The selected point is immediately after the first 3D convolutional layer that follows voxelization.
SC-MII adopts this splitting point to enable all subsequent processing without access to the original point cloud.

\subsection{Accuracy Evaluation of Integration Methods} \label{accuracy}

\begin{table}[t]
    \centering
    \caption{Overall Accuracy}
    \scalebox{0.95}{
    \begin{tabular}{llcc}
        \hline
        \multicolumn{1}{c}{\textbf{Sensor}} & \multicolumn{1}{c}{\multirow{2}{*}{\textbf{Integration methods}}} & \multicolumn{2}{c}{\textbf{mAP}} \\
        \multicolumn{1}{c}{\textbf{configuration}} & \ & \multicolumn{1}{c}{AP@0.3} & AP@0.5 \\
        \hline
        Single LiDAR & LiDAR 1 (No integration) & 71.79 & 48.88 \\
        \ & LiDAR 2 (No integration) & 73.24 & 50.47 \\ \hline
        Multiple LiDARs &Input point clouds & 78.17 & 56.41 \\
        \cline{2-4}
        \ & Intermediate outputs \textbf{(SC-MII)} & & \\
        \ & \hspace{5mm} Max value selection & 74.54 & 53.53 \\
        \ & \hspace{5mm} One convolutional layer & \ & \ \\
        \ & \hspace{10mm} Kernel size: 1 & 75.72 & 54.09 \\
        \ & \hspace{10mm} Kernel size: 3 & \textbf{77.12} & \textbf{55.32} \\
        \hline
    \end{tabular}
    }
    \vspace{-4mm}
    \label{tab-acc_all}
\end{table}

This section compares the accuracy of SC-MII, the proposed intermediate output integration method, against two baselines: using a single LiDAR and integrating input point clouds from multiple LiDARs.
The overall comparison is shown in Table~\ref{tab-acc_all}.
Compared to the single-LiDAR setting, using multiple LiDARs improves detection accuracy, possibly because multiple LiDARs provide denser points in distant regions and help recover object shapes that are partially occluded from a single viewpoint.
Among multi-LiDAR methods, SC-MII showed only a slight drop in accuracy compared to input point cloud integration, with reductions of 1.05\% at AP@0.3 and 1.09\% at AP@0.5.
While input integration benefits from access to raw data, SC-MII achieves comparable performance despite operating on intermediate outputs.

\subsection{Execution Time of Integration Methods} \label{execution_time}

This section compares the execution time of SC-MII with a baseline that performs all processing, from integrating input point clouds from multiple LiDARs to generating detection results, on a single edge device.

\begin{figure}[t]
        \centerline{\includegraphics[width=1.0\linewidth]{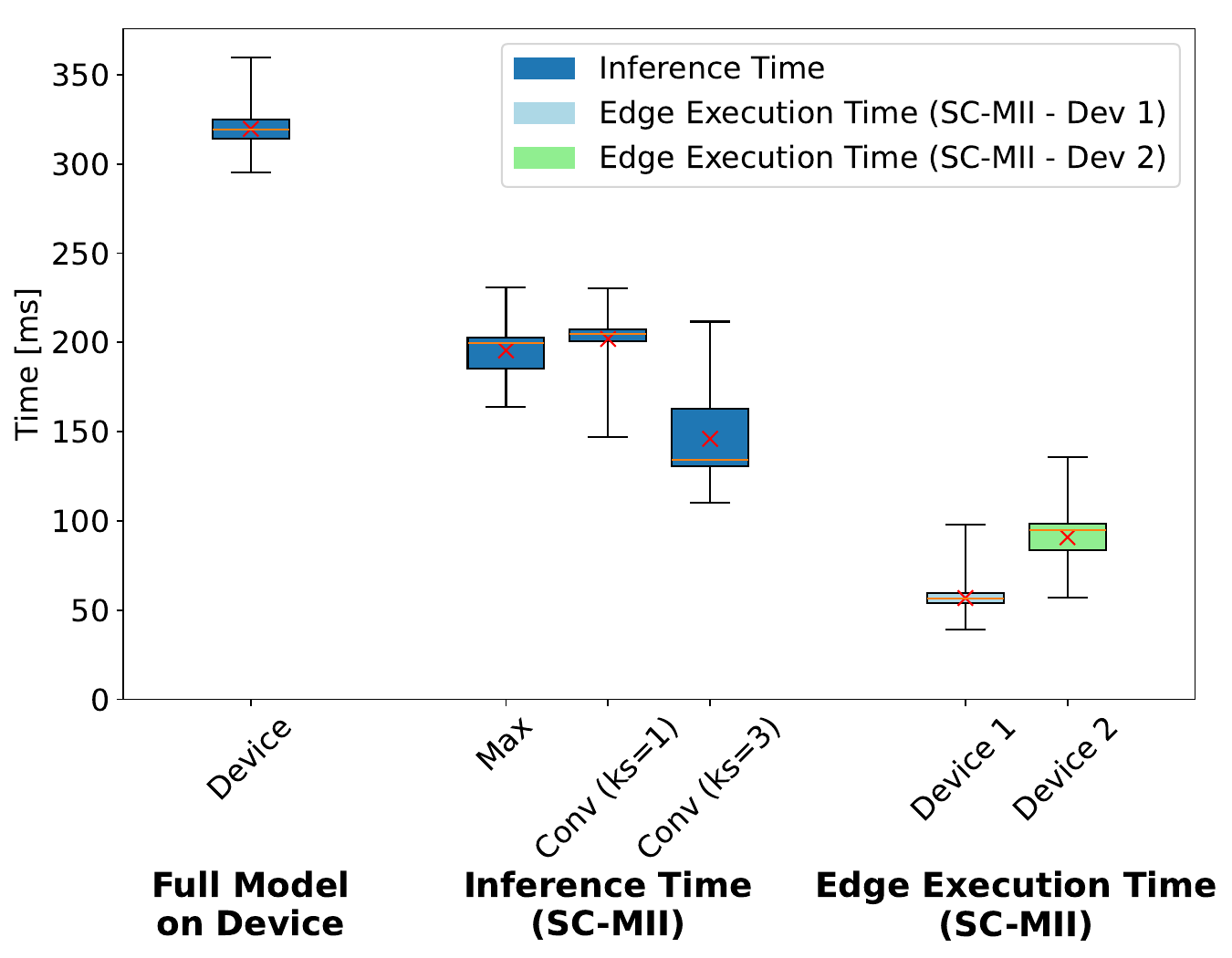}}
        \caption{Comparison of execution times.}
        \vspace{-5mm}
    \label{fig-execution_time}
\end{figure}

The inference time and the edge device execution time are shown in Fig.~\ref{fig-execution_time}.  
The inference time is the duration from point cloud input to final detection output.  
The edge device execution time is the time from input to the completion of intermediate output transmission under SC-MII.
Compared to full edge-side execution, SC-MII reduced inference time across all integration methods.
The method using a single convolutional layer with a kernel size of 3 showed the greatest speedup, with an average of 2.19× and a maximum of 1.70×.

Additionally, as all SC-MII variants share the same edge-side processing, the edge device execution time is presented per device.
In contrast, when all processing is performed on the edge device, the edge device execution time is equal to the overall inference time.
On Device 2, which handles a larger computational load, SC-MII reduced the edge-side processing time by an average of 71.6\% and a maximum of 62.2\%.

\subsection{Lessons Learned}

This section summarizes the key insights obtained from the evaluation results of this study.
These insights provide valuable implications for designing efficient and accurate distributed perception systems in autonomous driving systems.

SC-MII significantly improved the accuracy of 3D object detection by utilizing multiple infrastructure LiDARs, which helped reduce blind spots and increase point cloud density.
Compared to input-level integration, intermediate output integration resulted in only a slight drop in accuracy while maintaining comparable overall performance, demonstrating the effectiveness of the proposed approach.
Among the integration methods, using a single convolutional layer with a kernel size of 3 achieved the highest detection accuracy, outperforming both max value selection and convolution with a kernel size of 1.
This suggests that incorporating spatial context during integration is beneficial for accurate detection.

SC-MII significantly reduced processing time on resource-constrained edge devices by offloading computation to a high-performance edge server.  
Even though offloading introduces communication overhead, the overall inference latency was still reduced.  
This suggests that in scenarios where only low-power edge devices are available, delegating heavy computation to the edge server can effectively support real-time performance.  
Nevertheless, potential instability in network conditions highlights the need for further optimization of communication efficiency.

To improve communication robustness in SC, future work should explore techniques for reducing the size of data transmitted from edge devices.  
Integrating compressed intermediate outputs can help achieve a better trade-off between accuracy and latency.
Furthermore, systems designed to tolerate partial data loss without retransmission could enhance real-time performance in unstable network conditions.  
These directions can support the development of more reliable and efficient 3D perception systems for distributed environments.

\section{Related Work} \label{related_work}

\begin{table}[t]
    \centering
    \caption{Comparison of the Proposed Method with Related Work}
    \scalebox{0.94}[0.94]{
    \begin{tabular}{lccccc} \hline
        \ & \multirow{2}{*}{Edge} & \ & Multiple & \multirow{2}{*}{3D Object} & \multirow{2}{*}{Point} \\
        \ & \multirow{2}{*}{Device} & SC & Intermediate & \multirow{2}{*}{Detection} & \multirow{2}{*}{Cloud} \\
        \ & \ & \ & Integration & \ & \ \\ \hline
        Voxel R-CNN~\cite{deng2021voxel} & \ & \ & \ & \checkmark & \checkmark \\ \hline
        Lightweight model & \multirow{2}{*}{\checkmark} & \ & \ & \multirow{2}{*}{\checkmark} & \multirow{2}{*}{\checkmark} \\
        for 3D detection~\cite{li2023lightweight} & \ & \ & \ & \ & \ \\ \hline
        3D detection & \multirow{2}{*}{\checkmark} & \multirow{2}{*}{\checkmark} & \ & \multirow{2}{*}{\checkmark} & \multirow{2}{*}{\checkmark} \\
        with SC~\cite{noguchi20243d} & \ & \ & \ & \ & \ \\ \hline
        BottleFit~\cite{bottlefit} & \checkmark & \checkmark & \ & \ & \ \\ \hline
        ImVoxelNet~\cite{rukhovich2022imvoxelnet} & \ & \ & \checkmark & \checkmark & \ \\ \hline
        NNFacet~\cite{chen2023nnfacet} & \checkmark & \checkmark & \checkmark & \ & \ \\ \hline
        SC-MII (ours) & \checkmark & \checkmark & \checkmark & \checkmark & \checkmark \\ \hline
    \end{tabular}
    }
    \vspace{-4mm}
    \label{comparison}
\end{table}

This section reviews related work on SC and 3D object detection using multiple data sources.
The comparison results with the proposed method are summarized in Table~\ref{comparison}.

Voxel R-CNN~\cite{deng2021voxel} is a 3D object detection model that converts point cloud data into voxels to reduce computational cost.
A 3D Backbone extracts features from the voxels, which are then transformed into 2D features for region proposal generation using a 2D Backbone and RPN.
Voxel RoI Pooling, introduced in this model, enables efficient extraction of RoI features, achieving speed gains while maintaining accuracy.

A Lightweight Model for 3D Point Cloud Object Detection~\cite{li2023lightweight} targets edge devices with limited computational resources.
This model introduces the LW-Sconv module, combining factorized and group convolutions to reduce model complexity.
Knowledge distillation further improves accuracy, enabling efficient detection with minimal performance loss.

3D Point Cloud Object Detection on Edge Devices for Split Computing~\cite{noguchi20243d} applies SC to reduce the high computational cost of 3D object detection on resource-limited edge devices.
Offloading computation via SC lowers inference time and power consumption compared to full on-device execution.

BottleFit~\cite{bottlefit} introduces bottlenecks into DNNs for SC without increasing model complexity, unlike previous methods using autoencoders.
This method modifies the original model to reduce layers while enabling compression, and adopts a two-step training process for the head and tail models.
This approach improves accuracy while reducing communication latency and power consumption.

ImVoxelNet~\cite{rukhovich2022imvoxelnet} performs 3D object detection using arbitrary numbers of multi-view RGB images.
2D features from each image are projected into a unified 3D space and averaged per voxel to form a 3D feature map.
This map is processed by 3D convolutions and a detection head.
This achieves higher accuracy than prior RGB-based methods.

NNFacet~\cite{chen2023nnfacet} proposes a DNN-splitting framework for concurrent smart sensors, where edge devices extract class-specific features and transmit intermediate outputs to a server for fusion and classification.
The overall architecture closely resembles SC-MII, as both perform collaborative inference via feature sharing while reducing device load and preserving data locality.
However, SC-MII is tailored to 3D object detection with infrastructure LiDARs, introducing unique challenges in spatial alignment and feature integration.

\section{Conclusion} \label{conclusion}

This paper proposed SC-MII, a multiple infrastructure LiDAR-based 3D object detection method for edge devices, which aims to reduce inference time and computational load on the edge device.
SC-MII applies the concept of SC and enables object detection using multiple LiDARs by integrating multiple intermediate outputs.
The object detection model is split at a selected point, and the subsequent processing is offloaded to an edge server.
In this study, SC-MII was evaluated using Jetson Orin Nano as the edge device. Voxel R-CNN was employed as the object detection model.
Compared to edge-only processing that integrates input point clouds, SC-MII achieved shorter inference time and significantly reduced the computational load. The accuracy was maintained with only a slight degradation.

\bibliographystyle{IEEEtran}
\bibliography{reference}

\end{document}